\documentclass[aps,prl,preprint,nopacs,superscriptaddress]{revtex4-1}

\usepackage{graphicx}
\usepackage{verbatim}
\usepackage{mathrsfs}
\usepackage{color}
\usepackage{amsmath,amsfonts,amssymb}
\usepackage{graphicx}
\usepackage{color}
\usepackage{textcomp}
\usepackage{mathtools}
\usepackage{array}

\newcommand{\Z}{\mathbb{Z} }

\def\3{2.8in}    %used for figure widths
\def\2{2.5in}
\def\4{3.0in}

\def \beq {\begin{equation}}
\def \eeq {\end{equation}}
\begin{document}

\title{Purely Rotational Symmetry-Protected Topological Crystalline Insulator $\alpha$-Bi$_4$Br$_4$}

\author{Chuang-Han Hsu}
\affiliation{Department of Physics, National University of Singapore, Singapore 117542}
\affiliation{Centre for Advanced 2D Materials and Graphene Research Centre, National University of Singapore, Singapore 117546}

\author{Xiaoting Zhou}
\affiliation{Department of Physics, National Cheng Kung University, Tainan, 701, Taiwan}

\author{Qiong Ma}
\affiliation{Department of Physics, Massachusetts Institute of Technology, Cambridge, Massachusetts 02139, USA}

\author{Nuh Gedik}
\affiliation {Department of Physics, Massachusetts Institute of Technology, Cambridge, Massachusetts 02139, USA}

\author{Arun Bansil}
\affiliation{Department of Physics, Northeastern University, Boston, Massachusetts 02115, USA}

\author{Vitor M. Pereira}
\affiliation{Department of Physics, National University of Singapore, Singapore 117542}
\affiliation{Centre for Advanced 2D Materials and Graphene Research Centre, National University of Singapore, Singapore 117546}

\author{Hsin Lin}
%\email{nilnish@gmail.com}
\affiliation{Institute of Physics, Academia Sinica, Taipei 11529, Taiwan}

\author{Liang Fu}
\email{liangfu@mit.edu}
\affiliation {Department of Physics, Massachusetts Institute of Technology, Cambridge, Massachusetts 02139, USA}

\author{Su-Yang Xu}
\email{suyangxu@mit.edu}
\affiliation {Department of Physics, Massachusetts Institute of Technology, Cambridge, Massachusetts 02139, USA}

\author{Tay-Rong Chang}
\email{u32trc00@phys.ncku.edu.tw}
\affiliation{Department of Physics, National Cheng Kung University, Tainan, 701, Taiwan}
\affiliation{Center for quantum frontiers of research \& technology (QFort)}

\date{\today}

\begin{abstract}

Recent theoretical advances have proposed a new class of topological crystalline insulator (TCI) phases protected by rotational symmetries. Distinct from topological insulators (TIs), rotational symmetry-protected TCIs are expected to show unique topologically protected boundary modes: First, the surface normal to the rotational axis features  ``unpinned'' Dirac surface states whose Dirac points are located at generic $k$ points. Second, due to the ``higher-order'' bulk boundary correspondence, a 3D TCI also supports 1D helical edge states. Despite the unique topological electronic properties, to date, purely rotational symmetry-protected TCIs remain elusive in real materials. Using first-principles band calculations and theoretical modeling, we identify the van der Waals material $\alpha$-Bi$_4$Br$_4$ as a TCI purely protected by rotation symmetry. We show that the Bi$_4$Br$_4$'s $(010)$ surface exhibits a pair of unpinned topological Dirac fermions protected by the two-fold rotational axis. These unpinned Dirac fermions show an exotic spin texture highly favorable for spin transport and a band structure consisting of van Hove singularities due to Lifshitz transition. We also identify 1D topological hinge states along the edges of an $\alpha$-Bi$_4$Br$_4$ rod. We further discuss how the proposed topological electronic properties in $\alpha$-Bi$_4$Br$_4$ can be observed by various experimental techniques. 

\end{abstract}
\maketitle

\section{Introduction}

Topological crystalline insulators (TCIs) are insulators with nontrivial topology protected by crystalline symmetries
 \cite{hasan2010colloquium, Qi2011, bansil2016colloquium, fu2011topological}. While the large number of crystalline space group symmetries implies a rich variety of TCI states, for a long time, only mirror symmetry-protected TCIs were known \cite{teo2008surface, fu2011topological, hsieh2012topological, weng2014topological, wieder2017wallpaper, wang2016hourglass, tanaka2012experimental, dziawa2012topological, xu2012TCI, okada2013observation, zeljkovic2014mapping, liang2013evidence, li2016interfacial, chang2016discovery, sessi2016robust, liang2017pressure}. Recently, a new class of TCI states protected by $N$-fold rotational symmetries were proposed \cite{fang2017rotation}. Distinct from known topological insulating phases, rotational symmetry-protected TCIs are predicted to show ``unpinned'' Dirac surface states. Specifically, the surface normal to the rotational axis hosts $N$ Dirac cones whose the Dirac points appear at generic $k$ points in the surface Brillouin zone. Apart from the unpinned Dirac surface states, a 3D TCI also supports 1D topological edge states due to the ``higher order'' bulk-boundary correspondence \cite{song2017d, schindler2017higher, khalaf2018higher, matsugatani2018connecting, schindler2018higher, wang2018higher, yue2018symmetry}. In a rod with $N$-fold rotational symmetry that is finite sized along the two directions perpendicular to the rotational axis, the hinges of the rod host $N$ helical 1D edge states. Despite these unique topological boundary modes, purely rotational symmetry-protected TCIs remain elusive in real materials.
 
One challenge is that the rotational symmetry topological invariants are difficult to compute in first-principles calculations as they are defined in terms of Wannier center flow \cite{taherinejad2014wannier}. Recent theoretical studies have made significant progress on relating crystalline symmetry-protected topological invariants with symmetry eigenvalues of the electronic states \cite{bradlyn2017topological, kruthoff2017topological, po2017symmetry, song2017mapping, khalaf2017symmetry, fang2017diagnosis, bradlyn2018band, cano2018building}. In particular, Song et al \cite{song2017mapping} and Khalaf et al \cite{khalaf2017symmetry}, found that when certain additional symmetry $Y$ is present, topological invariants of TCIs protected by symmetry $X$ can be inferred by the $Y$-symmetry eigenvalues of energy bands. The proposal of symmetry indicators has facilitated first-principles studies of new TCIs \cite{zhou2018topological, tang2018efficient, tang2018topological, zhang2018catalogue, vergniory2018high}.

\begin{figure}
\begin{centering}
\includegraphics[scale=0.22]{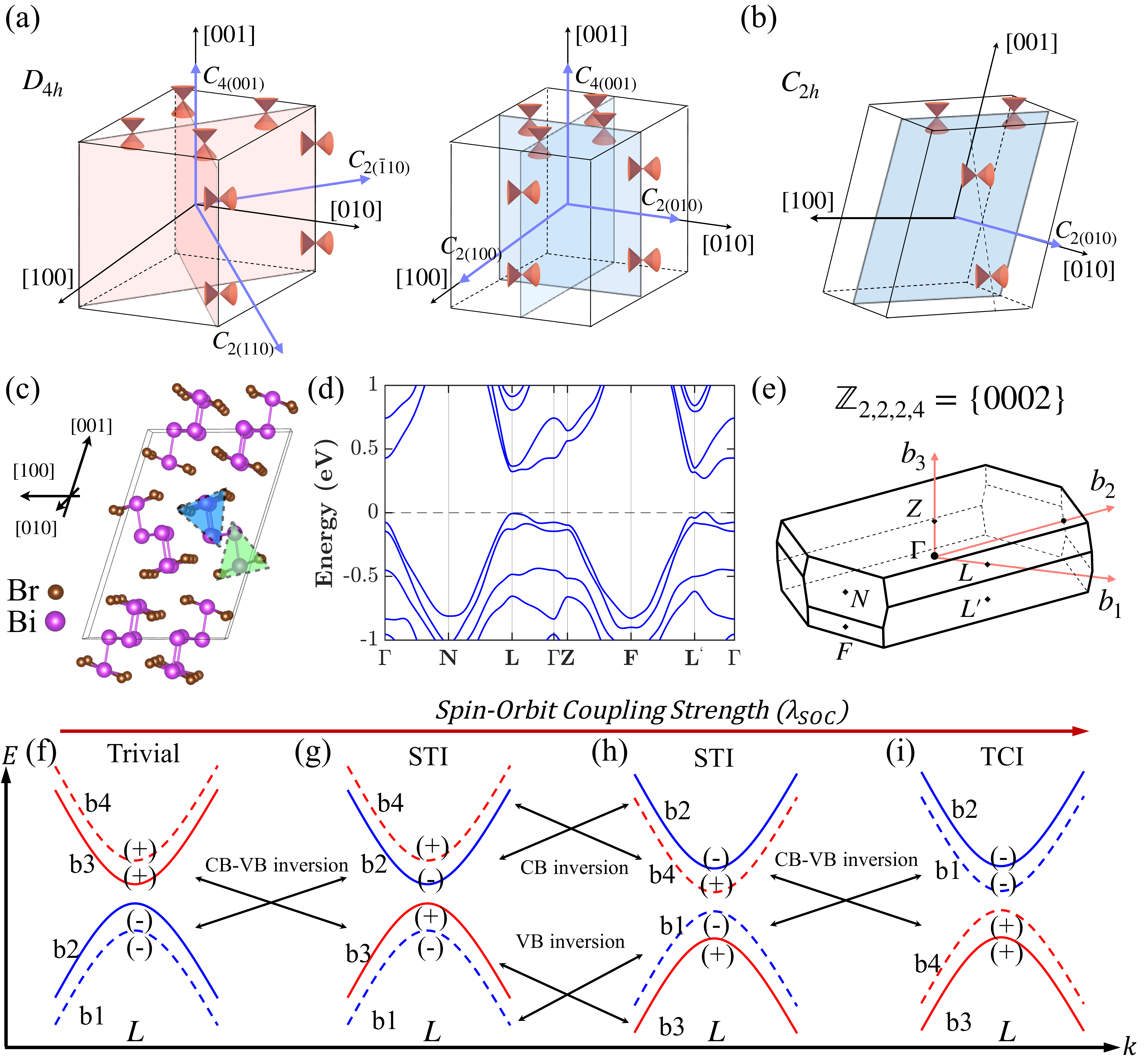}
\par\end{centering}
\centering{}\caption{(a) Schematic illustrations of TCI surface states on a tetragonal $D_{4h}$ lattice on different surfaces. In a $D_{4h}$ lattice, nontrivial topology protected by the out-of-plane four-fold rotational symmetry $4_{[001]}$ would lead to four Dirac fermions on the $(001)$ surface. However, because of nontrivial mirror planes ($M_{[100]}$,  $M_{[010]}$,  $M_{[110]}$ and  $M_{[1\bar{1}0]}$), the four Dirac fermions must be located along the $[100]$, $[010]$, $[110]$ or $[1\bar{1}0]$ high symmetry direction. (b) A $C_{2h}$ lattice consists of three symmetry operations, a two-fold rotational axis $2_{[010]}$, a mirror plane $M_{[010]}$, and space inversion symmetry $\mathcal{I}$. The $(010)$ surface only hosts the two-fold rotational symmetry $2_{[010]}$. Therefore, $C_{2h}$ crystals are ideal for searching for purely rotational symmetry-protected TCIs and the two ``unpinned'' Dirac fermions can appear at generic $k$ points on the $(010)$ surface due to the nontrivial topology protected by $2_{[010]}$. (c) Crystal structure of $\alpha$-Bi$_4$Br$_4$. (d) Band structure with SOC. (e) Bulk Brillouin zone (BZ) of $\alpha$-Bi$_4$Br$_4$ with the high symmetry points noted. Crystals in space group $\#12$ (including $\alpha$-Bi$_4$Br$_4$) are characterized by four symmetry indicators, three $\mathbb{Z}_2$ and one $\mathbb{Z}_4$ ($\mathbb{Z}_{2,2,2,4}$) \cite{song2017mapping, khalaf2017symmetry}. For $\alpha$-Bi$_4$Br$_4$, our calculations show that $\mathbb{Z}_{2,2,2,4}=\lbrace0002\rbrace$. (f-i) Schematic illustration of the evolution of the low-energy band structures as a function of the strength of SOC near the $L$ point. The parity eigenvalues of the electron states at $L$ are labeled. }
\label{Fig1}
\end{figure}

Another challenge is to isolate the nontrivial topology protected by rotational symmetry from that of protected by mirror symmetry. Taking the example of a tetragonal crystal lattice (point group $D_{4h}$, space group $P4/mmm$, Fig.~\ref{Fig1}(a)). Nontrivial topology protected by the out-of-plane four-fold rotational symmetry $4_{[001]}$ would lead to four Dirac fermions on the $(001)$ surface (Fig.~\ref{Fig1}(a)). The conventional basis is used for $[hkl]$ and $(hkl)$ here (see supplementary materials (SM)). However, the band topology gives rise to the nontrivial $4_{[001]}$ index is likely to harbor the mirror symmetry-protected nontrivial topology in mirror planes ($M_{[100]}$,  $M_{[010]}$,  $M_{[110]}$ and  $M_{[1-10]}$). Consequently, the four Dirac fermions must be located along the $[100]$, $[010]$, $[110]$ or $[1\bar{1}0]$ high symmetry direction (Fig.~\ref{Fig1}(a)). We propose that a particular class of monoclinic lattice with the point group $C_{2h}$ is ideal for searching for purely rotational symmetry-protected TCIs. As shown in Fig.~\ref{Fig1}(b), a $C_{2h}$ lattice consists of three symmetry operations, a two-fold rotational axis $2_{[010]}$, a mirror plane $M_{[010]}$, and space inversion symmetry $\mathcal{I}$. Interestingly, the $(010)$ surface only preserves the two-fold rotational symmetry $2_{[010]}$ but breaks $M_{[010]}$. On the other hand, the presence of  inversion symmetry $\mathcal{I}$ in the bulk still allows the application of the symmetry indicator theory. Therefore, we propose that $C_{2h}$ crystals are ideal for searching for purely rotational symmetry-protected TCIs and the two ``unpinned'' Dirac fermions can appear at generic $k$ points on the $(010)$ surface due to the nontrivial topology protected by $2_{[010]}$.

Guided by the analyses that we propose above, we have carefully searched real materials with a monoclinic $C_{2h}$ lattice. In this paper, we have identified the van der Waals material $\alpha$-Bi$_4$Br$_4$ as a TCI purely protected by rotation symmetry. Figure~\ref{Fig1}(c) depicts the experimentally determined lattice structure of $\alpha$-Bi$_4$Br$_4$ (point group $C_{2h}$, space group $C2/m$ ($\#12$)) \cite{von1978kenntnis, filatova2007electronic}. Consistent with the general picture in Fig.~\ref{Fig1}(b), the two-fold rotational axis of the $\alpha$-Bi$_4$Br$_4$ lattice is along the $[010]$ direction (Fig.~\ref{Fig1}(c)). We note that recent experiments \cite{Aut2016, pisoni2017pressure, qi2018pressure} have identified a topological insulator state in the sister compound $\beta$-Bi$_4$I$_4$, which has a slightly different crystal structure (see SM). On the other hand, the electronic and topological properties of $\alpha$-Bi$_4$Br$_4$ remains largely unexplored, except that its monolayer has been proposed to be a 2D quantum spin Hall insulator \cite{zhou2014large}.

\section{Crystal Structure and Band Topology}

We now study the electronic structure of $\alpha$-Bi$_4$Br$_4$ via first-principles methods. The band structure calculations were carried out by the VASP package \cite{DFT2} and the experimental crystal structure of $\alpha$-Bi$_4$Br$_4$ was used \cite{von1978kenntnis}. The Wannier basis  \cite{souza2001maximally} based tight-binding model was extrapolated for studying surface states of $\alpha$-Bi$_4$Br$_4$. The band structure with spin-orbit coupling (SOC) (Fig.~\ref{Fig1}(d)) shows a band gap around $0.4$ eV. The bands at low energies are dominated by four bands, two conduction bands (CBs) and two valence bands (VBs), all of which are mainly from the Bi $6p$-orbitals. By examining the evolution of the band structure as a function of SOC strength, we have identified four band inversions between bands of opposite parity eigenvalues at the $L$ point (Figs.~\ref{Fig1}(f-i)). Since the bulk Brillouin zone (Fig.~\ref{Fig1}(e)) consists of only one $L$ point, the even number of band inversions rules out the possibility for a time-reversal symmetry-protected $Z_2$ TI state.

\begin{table*}
\centering
\begin{tabular}{>{\centering\arraybackslash}m{4cm}>{\centering\arraybackslash}m{4cm}>{\centering\arraybackslash}m{4cm}}
\hline
$(\nu_0;\nu_1\nu_2\nu_3)$ & $n_{2_{[010]}}$ & $n_{\mathcal{M}_{[010]}}$\\
\hline
(0;000)& 1 & 0 \\
(0;000)& 0 & 2 \\
\hline
\end{tabular}
\caption{The two possible topological states for the symmetry indicator $\Z_{2,2,2,4} =\lbrace 0,0,0,2 \rbrace$ according to Ref. \cite{song2017mapping}. $(\nu_0;\nu_1\nu_2\nu_3)$ are the $\mathbb{Z}_2$ invariants for 3D $\mathbb{Z}_2$ TIs. $\nu_0=1$ corresponds to a strong TI; $\nu_0=0$ but $\nu_1+\nu_2+\nu_3\neq0$ corresponds to a weak TI. $n_{2_{[010]}}$ is the topological invariant for the two-fold rotational symmetry $2_{[010]}$, and is also a $\mathbb{Z}_2$ invariant. $n_{2_{[010]}}=1$ corresponds to a rotational symmetry-protected TCI, which features two ``unpinned'' Dirac surface states on the $(010)$ surface. $n_{\mathcal{M}_{[010]}}$ is the topological invariant for mirror plane $\mathcal{M}_{[010]}$ (the mirror Chern number), and is a $\Z_N$ invariant. $n_{\mathcal{M}_{[010]}}=N$ corresponds to a mirror symmetry-protected TCI, which features $N$ Dirac surface states. See SM for detailed introduction to the symmetry indicators and the topological invariants. }
\label{Tab1}
\end{table*}

We then calculate the symmetry-based indicators (SIs). As shown by Refs. \cite{song2017mapping, khalaf2017symmetry}, crystals in space group $\#12$ are characterized by four symmetry indicators, three $\Z_2$ and one $\Z_4$ ($\Z_{2,2,2,4}$). By enumerating the symmetry eigenvalues of the electron states at high symmetry points, we obtain $\Z_{2,2,2,4} =\lbrace 0,0,0,2 \rbrace$ for $\alpha$-Bi$_4$Br$_4$. Table~\ref{Tab1} shows the topological states corresponding to $\Z_{2,2,2,4} =\lbrace 0,0,0,2 \rbrace$ according to Ref. \cite{song2017mapping} (see the caption of Tab.~\ref{Tab1} for the definition of the topological invariants). Importantly, the specific symmetry indicators ($\Z_{2,2,2,4} =\lbrace 0,0,0,2 \rbrace$) point to two possible topological states, one being a purely rotational symmetry-protected TCI with $n_{2_{[010]}}=1$ and the other one being a purely mirror symmetry-protected TCI with $n_{\mathcal{M}_{[010]}}=2$. Note that the $\mathcal{I}$-protected topology is referred to the higher order topological insulator (HOTI)~\cite{schindler2018higher}, and thus is not considered here. In order to uniquely determine the topological state of $\alpha$-Bi$_4$Br$_4$, we have further calculated the mirror Chern number $n_{\mathcal{M}_{[010]}}$ and found $n_{\mathcal{M}_{[010]}}=0$ for $\alpha$-Bi$_4$Br$_4$. Therefore, these calculations reveal that $\alpha$-Bi$_4$Br$_4$ is a purely rotational symmetry-protected TCI with $n_{2_{[010]}}=1$. 

\section{Surface states}

\begin{figure}
\begin{centering}
\includegraphics[scale=0.3]{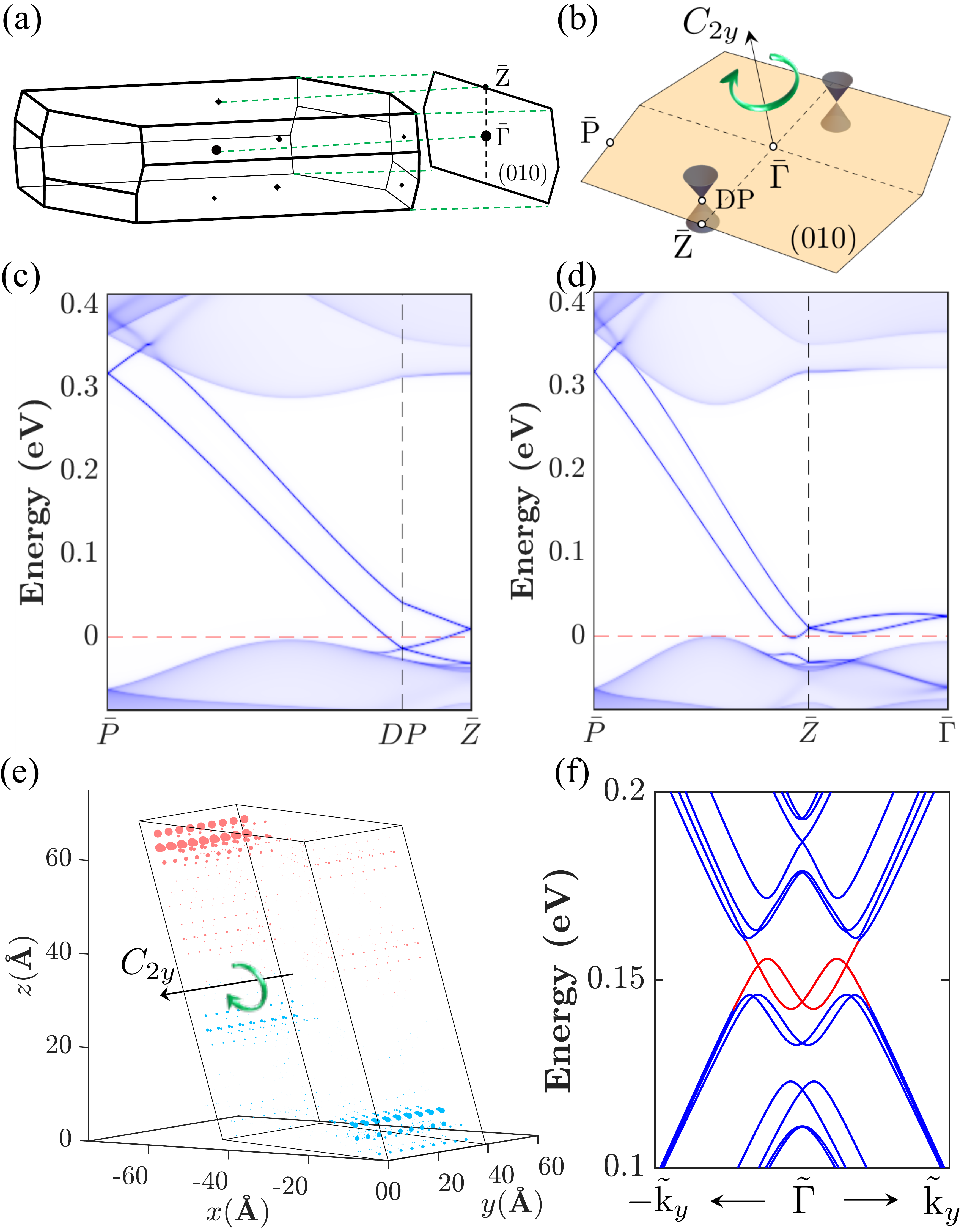}
\par\end{centering}
\centering{}\caption{(a) Bulk BZ and the $(010)$ surface BZ. (b) Schematic illustration for the unpinned topological Dirac cones protected by the two-fold rotational symmetry $2_{[010]}$ on the $(010)$ surface. (c)(d) Calculated $(010)$ surface states along $\bar{P}-\bar{\textrm{DP}}-\bar{Z}$ and $\bar{P}-\bar{Z}-\bar{\Gamma}$. (f) The band structure of 1D $\alpha$-Bi$_4$Br$_4$ rod that is finite sized along $[\bar{1}10]$ and $[001]$ but periodic along $[010]$. (e) Real space distribution of the nontrivial edge state indicated by the red curves shown in (f). 
%The black box shows the size of the 1D structure in real space and 
The arrow shows the direction along which the rod is periodic.}
\label{Fig2}
\end{figure}

To confirm the topological nature of $\alpha$-Bi$_4$Br$_4$, we study its novel topological boundary modes. We first focus on uncovering the ``unpinned'' Dirac surface states on the $(010)$ surface protected by the two-fold rotational symmetry $2_{[010]}$. By calculating the band structure throughout the $(010)$ surface BZ, we have indeed identified two Dirac cones. Their Dirac points (DPs)are located at $\pm(0.0434\;\frac{2\pi}{a},0.3550\;\frac{2\pi}{c})$. These are generic $k$ points only related by the two-fold rotational symmetry $2_{[010]}$. The calculated surface states along a $k$ path passing the DP ($\bar{P}-\bar{\textrm{DP}}-\bar{Z}$, Fig.\ref{Fig2}(c)) directly shows the topological Dirac surface states. On the other hand, along the path $\bar{P}-\bar{Z}-\bar{\Gamma}$ that does not go through the DP, the surface states are gapped (Fig.\ref{Fig2}(d)). We now demonstrate the 1D topological hinge states protected by $2_{[010]}$. We have calculated the band structure of 1D $\alpha$-Bi$_4$Br$_4$ rod that is finite sized along $[\bar{1}10]$ and $[001]$ but periodic along $[010]$. In Fig.\ref{Fig2}(f) we show the existence of 1D helical edge states inside the bulk band gap. We then study the real space distribution of the electron wave function for the 1D helical edge states. As shown in Fig.\ref{Fig2}(e), these helical states are localized on the edges shared by adjacent side surfaces, confirming that they are topological hinge states. Experimentally, the ``unpinned'' Dirac surface states can be directly imaged by photoemission spectroscopy, whereas the helical ``hinge'' states may be probed by scanning tunneling spectroscopy on a nanorod.

\begin{figure}
\begin{centering}
\includegraphics[scale=0.4]{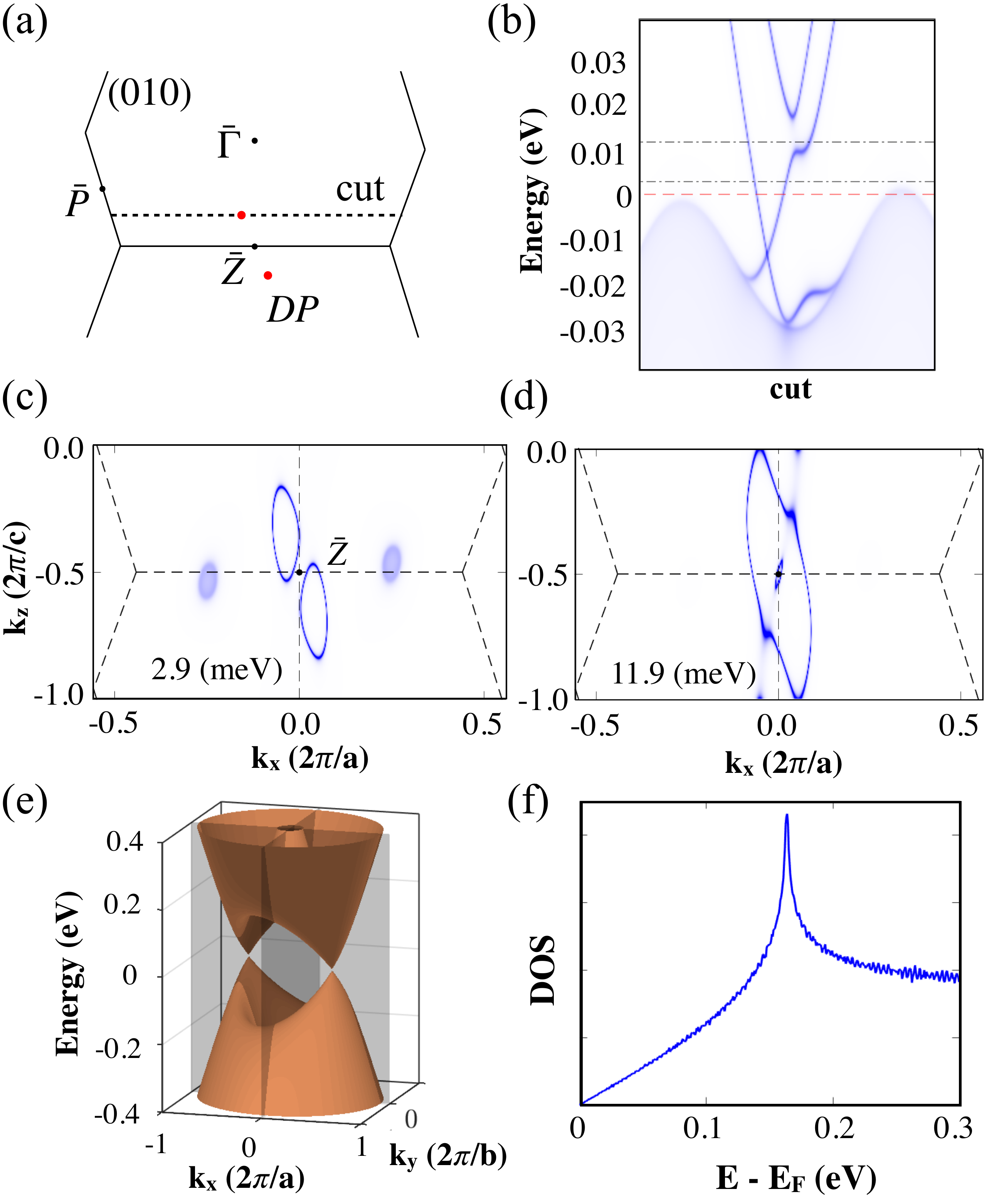}
\par\end{centering}
\centering{}\caption{(a) The $(010)$ surface BZ. The red points denote the center of the unpinned Dirac cones. (b) The surface state band structure calculated along the path defined by the black dotted line in (a). (c,d) Constant energy contours at the binding energies of 2.9 meV and 11.9 meV (indicated in panel (b) by the dot lines). A Lifshitz transition is observed as two separate contours (panel (c)) merge and become two concentric contours (panel (d)). (e,f) Electronic structure generated by the $k\cdot p$ model that describes a pair of ``unpinned'' Dirac cones on the $(010)$ surface of $\alpha$-Bi$_4$Br$_4$. (f) Calculated density of states (DOS) as a function of energy based on the $k\cdot p$ model. A sharp DOS peak is observed at the energy corresponding to the Lifshitz transition, demonstrating the van Hove singularity. }
\label{Fig3}
\end{figure}

We further highlight novel electronic properties enabled by the ``unpinned'' topological Dirac fermions on the $(010)$ surface. First, we show the existence of van Hove singularities on the $(010)$ surface. At an energy close to the Dirac point ($E=2.9$ meV), the surface state exhibits two separated electron-like contours (Fig.~\ref{Fig3}(c)). As one increases the energy away above the Dirac point, the contours expand and eventually merge. At $E=11.9$ meV, two concentric contours are realized (Fig.~\ref{Fig3}(d)). Such a Lifshitz transition signals a saddle point in the surface electronic band structure, resulting in a van Hove singularity (VHS) with diverging density of states (DOS). To gain more insights to the ``unpinned'' Dirac surface states and to uncover the VHS, we have constructed an effective $k\cdot p$ model. The $(010)$ surface has two symmetries, the two fold rotation $2_{[010]}$ and time-reversal $\mathcal{T}$. Based on the first-principles calculations, we have chosen the following symmetry adapted bases for the $k\cdot p$ model. $|\Psi_{\mathbf{k}} \rangle = {(\psi_{+\uparrow}, \psi_{+\downarrow} , \psi_{-\uparrow}, \psi_{-\downarrow} )}^{T}$ as,
\begin{eqnarray}\label{eq:bases}
|\psi_{\pm,s} \rangle &=& \sum_{\alpha = x,z} \lambda_{\alpha}s (| p^A_{\alpha}, s \rangle \pm | p^B_{\alpha}, s \rangle) \nonumber \\
&& + \;\lambda_{y}s (| p^A_{y}, s \rangle \mp | p^B_{y}, s \rangle),
\end{eqnarray} where $\pm$ is the eigenvalue of $2_{[010]}\times\mathcal{T}$, $s\;(=\uparrow\downarrow)$ is the spin index, and $\lambda_{i}$ is the orbital weight for the $p_{i}$ orbit. The A and B sublattices are indicated by blue and green triangular in Fig.~\ref{Fig1}(c). The (010) surface on the Cartesian coordinate is spanned on the $x$-$z$ plane.  The simplest $4 \times 4$ dimensional model including all the allowed constant terms $\delta_{ij}$ takes the form as,
\begin{eqnarray}\label{eq:Hamiltonian}
H(\mathbf{k}) &=&v_R\tau_{0} (k_{x} \sigma_{1} - k_{z} \sigma_{3}) + \delta_{30} \tau_3 \sigma_0 + \delta_{21} \tau_2 \sigma_1 \nonumber  \\
&& +\; \delta_{23} \tau_2 \sigma_3,
\end{eqnarray} where the Pauli matrices $\sigma_i$ and $\tau_i$ denote the spin and sublattice degree of freedom, respectively.
As shown in Figs.~\ref{Fig3}(e,f), the Lifshitz transition is reproduced by the $k\cdot p$ model. We calculate the DOS as a function of energy based on the $k\cdot p$ model. A divergent behavior at the energy position of the VHS is clearly visualized (Fig.~\ref{Fig3}(f)). The VHS and the corresponding diverging DOS are favorable conditions for correlated quantum phenomena \cite{li2010observation, nilsson2006electron, nandkishore2012chiral}. 

\begin{figure}
\begin{centering}
\includegraphics[scale=0.4]{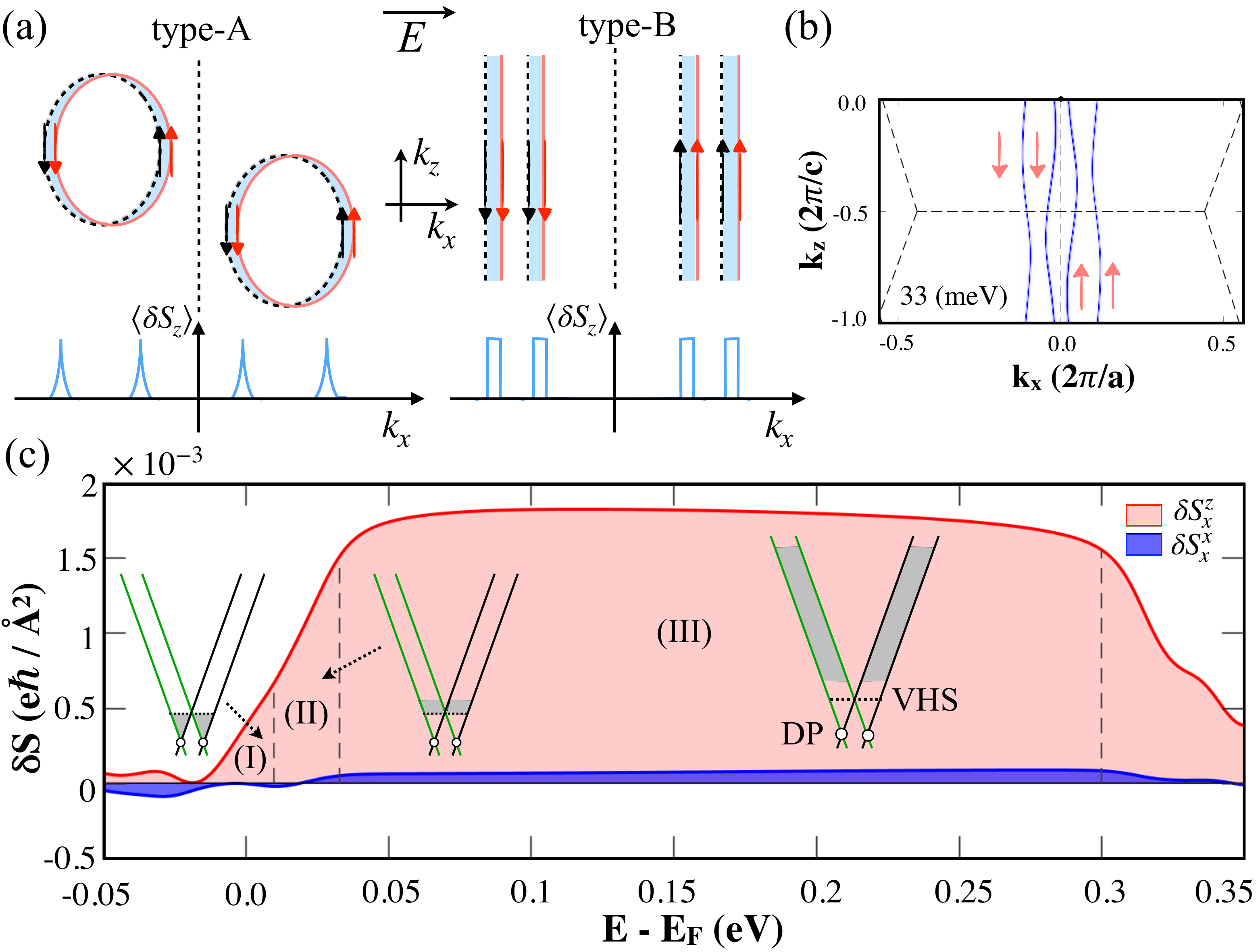}
\par\end{centering}
\centering{}\caption{(a) Schematic picture illustrates the nonequilibrium spin density ($\delta S$) driven by the external electric field (E) for the type-A (left) Fermi surface, that has been shown in Fig.~\ref{Fig3}(c), and the type-B (right), that is shown in (b) for $\mu$ = 33 meV on $\alpha$-Bi$_4$Br$_4$ (010) surface. The black dashed lines are the Fermi contours at equilibrium, and the red lines are the Fermi contours under the E field. The blue shaded areas indicate the induced $\delta S$ due to the Fermi surface shift. (c) Calculated $\delta S$ on the $\alpha$-Bi$_4$Br$_4$ (010) surfaces. $\delta S$ in the area (I) and (III) are given dominantly by the type-A and type-B in (a), respectively. The energy windows for (I) and (III) are indicated by the gray areas on the schematic E-k diagrams. }
\label{Fig4}
\end{figure}

Second, we study the spin texture of the ``unpinned'' Dirac fermions and discuss their connections to spin transport. At energies close the surface Dirac points (e.g. Fig.~\ref{Fig3}(c)), the spin texture is quite similar to that of existing topological insulators: Spin polarization winds tangentially around the closed constant energy contours. On the other hand, as one increases the energy further away from the Dirac point (e.g. $E=33$ meV in Fig.~\ref{Fig4}(b)), the constant energy contour evolves into separated lines that transverse the whole surface BZ. Interestingly, in this regime, the two left movers are spin-up whereas the two right movers are spin-down, realizing a 3D quantum spin Hall effect \cite{wang2016hourglass}. Importantly, in $\alpha$-Bi$_4$Br$_4$, this spin texture is realized in a large energy window (0 meV - 300 meV) due to the large bulk energy gap. We propose that such a unique spin texture can lead to large spin transport effect. Here, we calculate the current-induced spin accumulation ($\delta S$) due to the Edelstein effect \cite{edelstein1990spin}: Essentially, passing an electrical current tilts the Fermi surface out of equilibrium distribution, which can lead to a net spin polarization. Considering the unique spin texture, we expect a sizeable accumulation of the spin polarization along $z$ ($S^{z}$) when passing an electrical current along the $x$ direction. Theoretically, the spin accumulation can be calculated within the linear response theory \cite{li2015intraband} by,
\begin{equation}
\delta\mathbf{S}=\frac{e\hbar}{\Gamma V}\sum_n \delta(\varepsilon_{\textbf{F}}-\varepsilon_{n}) \langle n |\mathbf{S}|n \rangle \langle n |\mathbf{v}\cdot\mathbf{E}|n \rangle,
\end{equation} where $e$, $\hbar$, $V$ and $\Gamma$ are the unit charge, the Plank constant, volume of the unit cell, and broadening due to impurities, respectively, $\varepsilon_{\textbf{F}}$ is the Fermi energy, $\varepsilon_{n}$ are the energy eigenvalue of state $|n \rangle$, and $\mathbf{S}$, $\mathbf{v}=\frac{\partial H(\mathbf{k})}{\hbar\partial \mathbf{k}}$ and $\mathbf{E}$ are the spin, velocity and the external electric field. The calculated results are shown in Fig.~\ref{Fig4}(c), where $\delta S^{\alpha}_{\beta}$ denotes the spin polarization along the $\alpha$ direction by flowing an external current along the $\beta$ direction. The largest response is indeed $\delta S_{x}^{z}$. In regime (I), the surface states (SSs) and the spin texture resemble two copies of TI centered at generic momenta (referred as type-A behavior, Fig.~\ref{Fig4}(a)). $\delta S_{x}^{z}$ increases with increasing energy since the two Fermi contours enlarge. For regime (III), the four parallel Fermi contours with high $s_{z}$ polarization lead to largest $\delta S_{x}^{z}$, which is almost constant in this regime due to its quasi-1D characteristic (referred as type-B behavior, Fig~\ref{Fig4}(a)). Regime II is a transition region that connects regimes I and III. We expect that this unique spin accumulation can be directly probed by Kerr rotation microscopy~\cite{Kato2004a}. 

\section{Conclusion}

In conclusion, we propose $\alpha$-Bi$_4$Br$_4$ as a candidate of the newly proposed $C_{2}$ protected TCI. This TCI phase is confirmed by the unpinned Dirac surface states on the $(010)$ surface and the 1D topological hinge states on a $C_{2}$ preserving $\alpha$-Bi$_4$Br$_4$ rod. In addition, the (010) surface states at higher binding energies realize a 3D double quantum spin Hall effect, which supports large spin transport effect. %In the end, we also construct a surface $k.p$ Hamiltonian for the 2D SSs to simplify the study of the unpinned Dirac surface states. 

\textit{Note Added}: While completing this manuscript, we noticed Ref. \cite{tang2018efficient}, which also proposed $\alpha$-Bi$_4$Br$_4$ as a rotation-symmetry protected TCI.

%\suppinfo

{\bf Supporting Information Available}

%\begin{thebibliography}{99}

%\end{thebibliography}

%\bibliography{Topological_and_2D_05102018}

%\bibliographystyle{achemso}

\end{document}